\documentclass[twocolumn,showpacs,aps,prl,superscriptaddress,floatfix,letter]{revtex4}

\usepackage{epsfig}
\usepackage{graphics}
\usepackage{graphicx}
\usepackage{epic}
\usepackage{floatflt}
\usepackage{dcolumn}
\usepackage{amsmath}

\newcommand{\dkpi}{\ensuremath{D^\mp K^0 \pi^\pm}}

\newcommand{\bdkpi}{\ensuremath{B^0 \to D^\mp K^0 \pi^\pm}}
\newcommand{\bdkstar}{\ensuremath{B^0 \to D^\mp K^{*\pm}}}

\newcommand{\dskpi}{\ensuremath{D^{*\mp} K^0 \pi^\pm}}

\newcommand{\bdskpi}{\ensuremath{B^0 \to D^{*\mp} K^0 \pi^\pm}}
\newcommand{\bdskstar}{\ensuremath{B^0 \to D^{*\mp} K^{*\pm}}}

\newcommand{\bdpskpi}{\ensuremath{B^0 \to D^{(*)\mp} K^0 \pi^\pm}}
\newcommand{\bdpskstar}{\ensuremath{B^0 \to D^{(*)\mp} K^{*\pm}}}

\newcommand{\de}{\ensuremath{\mbox{$\Delta E$}}}
\newcommand{\fis}{\ensuremath{\mbox{$\mathcal{F}$}}}
\newcommand{\Bbar}{\ensuremath{\overline{B} \rule{0ex}{1ex}^0}}

\long\def\inst#1{\par\nobreak\kern 4pt\nobreak
    {\it #1}\par\vskip 10pt plus 3pt minus 3pt}

\def\sss{\scriptscriptstyle}
\def\barpd{{\raise.35ex\hbox{${\sss (}$}}--{\raise.35ex\hbox{${\sss )}$}}}
\def\dbarp{\hbox{$D^{0}$\kern-1.3em\raise1.5ex\hbox{\barpd}}}
\def\dbarpnozero{\hbox{$D$\kern-0.85em\raise1.5ex\hbox{\barpd}}}

\RequirePackage{xspace}

\usepackage{relsize}
\def\babar{\mbox{\slshape B\kern-0.1em{\smaller A}\kern-0.1em
    B\kern-0.1em{\smaller A\kern-0.2em R}}}

\def\Bbar  {\kern 0.18em\overline{\kern -0.18em B}{}\xspace}

\mathchardef\Upsilon="7107
\def\Y#1S{\ensuremath{\Upsilon{(#1S)}}\xspace}
\def\FourS {\Y4S}

\def\invfb   {\ensuremath{\mbox{\,fb}^{-1}}\xspace}
\def\KS    {\ensuremath{K^0_{\scriptscriptstyle S}}\xspace} 
\def\mes        {\mbox{$m_{\rm ES}$}\xspace}

\newcommand{\gev}{\ensuremath{\mathrm{\,Ge\kern -0.1em V}}\xspace}
\newcommand{\mev}{\ensuremath{\mathrm{\,Me\kern -0.1em V}}\xspace}
\newcommand{\gevc}{\ensuremath{{\mathrm{\,Ge\kern -0.1em V\!/}c}}\xspace}
\newcommand{\mevc}{\ensuremath{{\mathrm{\,Me\kern -0.1em V\!/}c}}\xspace}
\newcommand{\gevcc}{\ensuremath{{\mathrm{\,Ge\kern -0.1em V\!/}c^2}}\xspace}
\newcommand{\mevcc}{\ensuremath{{\mathrm{\,Me\kern -0.1em V\!/}c^2}}\xspace}
\def\to                 {\ensuremath{\rightarrow}\xspace}

\def\pep2{PEP-II}

\begin{document}
\begin{flushleft}
SLAC-PUB-10893 \\
\babar-PUB-04/40 
\end{flushleft}

\title{Measurements of Branching Fractions and Dalitz Distributions for
             {\boldmath $B^0 \to D^{(*)\pm} K^0 \pi^\mp$} Decays}
\author{B.~Aubert}
\author{R.~Barate}
\author{D.~Boutigny}
\author{F.~Couderc}
\author{Y.~Karyotakis}
\author{J.~P.~Lees}
\author{V.~Poireau}
\author{V.~Tisserand}
\author{A.~Zghiche}
\affiliation{Laboratoire de Physique des Particules, F-74941 Annecy-le-Vieux, France }
\author{E.~Grauges-Pous}
\affiliation{Universitad Autonoma de Barcelona, E-08193 Bellaterra, Barcelona, Spain }
% ...
\collaboration{The \babar\ Collaboration}
\noaffiliation

\date{\today}

\begin{abstract}
\noindent
We present measurements of the branching fractions for the three-body decays
\bdpskpi\ and their resonant submodes \bdpskstar\ using a sample of
approximately 88 million $B \Bbar$ pairs collected by the \babar\
detector at the PEP-II asymmetric energy storage ring.
We measure:
\begin{eqnarray*}
  {\cal B}(\bdkpi)    \!\! &=& \!\! 
  (4.9 \pm 0.7_{\, \mbox{\footnotesize stat}}  
        \pm 0.5_{\, \mbox{\footnotesize syst}}) \! \times \! 10^{-4},\\
  {\cal B}(\bdskpi)   \!\! &=& \!\! 
  (3.0 \pm 0.7_{\, \mbox{\footnotesize stat}}  
        \pm 0.3_{\, \mbox{\footnotesize syst}}) \! \times \! 10^{-4},\\
  {\cal B}(\bdkstar)  \!\! &=& \!\! 
  (4.6 \pm 0.6_{\, \mbox{\footnotesize stat}}  
        \pm 0.5_{\, \mbox{\footnotesize syst}}) \! \times \! 10^{-4},\\
  {\cal B}(\bdskstar) \!\! &=& \!\! 
  (3.2 \pm 0.6_{\, \mbox{\footnotesize stat}}  
        \pm 0.3_{\, \mbox{\footnotesize syst}}) \! \times \! 10^{-4}.
\end{eqnarray*}
From these measurements we determine the fractions of resonant events
to be
$f(\bdkstar)  = 0.63 \pm 0.08_{\,\mbox{\footnotesize stat}}
                     \pm 0.04_{\,\mbox{\footnotesize syst}}$ and
$f(\bdskstar) = 0.72 \pm 0.14_{\,\mbox{\footnotesize stat}}
                     \pm 0.05_{\,\mbox{\footnotesize syst}}$.
\end{abstract}


\maketitle


Several independent measurements are needed to test the Standard Model
description of $CP$ violation. The angle $\gamma$ can be determined using
decays of the type $B \to D^{(*)} K^{(*)}$ \cite{gwl}.
The experimental challenges are color suppression of the $b \to u$ transition,
reconstruction of $D^0$ $CP$ eigenstates, and interfering
doubly-Cabibbo-suppressed decays (DCSD) \cite{ads}. Also, two-body mode
analyses are complicated because there are eight degenerate solutions for
$\gamma$ in the interval $[0,2\pi]$.

In recent papers \cite{aps,ap} three-body decays have been suggested for
measuring $\gamma$, since these do not suffer from the color suppression
penalty. Furthermore, the channels \bdpskpi\ do not have the above
problems with $CP$ states and DCSD interference, and can resolve most of
the ambiguities \cite{aps}.
The angle $\gamma$ can be extracted from a time-dependent Dalitz analysis
of these decay modes.

The analysis presented here is based on 81.8 \invfb of data taken at the
$\Upsilon(4S)$ resonance, corresponding to approximately 88 million $B
\Bbar$ pairs, with the \babar\ detector \cite{babar-nim} at the PEP-II
storage ring.
We measure the branching fractions of the \bdpskpi\ decays and consider
their distribution in the Dalitz plot.

We reconstruct $D^+$ mesons in the decay mode $K^- \pi^+ \pi^+$
and $D^{*+}$ mesons in the mode $D^0 \pi^+$, with the $D^0$ decaying
to $K^- \pi^+$, $K^- \pi^+ \pi^0$, and $K^- \pi^+ \pi^- \pi^+$. Here
and throughout the paper charge conjugate states are implied.
Tracks from the $D$ decay are required to originate from a common vertex.
Positive kaon identification is enforced on kaons from $D$ meson
decays, except for the $D^0 \to K^- \pi^+$ mode.\\
\indent
The $D^+$ candidates are required to have a mass within 12 \mevcc\
(2$\sigma$) of the $D^+$ mass, while the mass of $D^0$ candidates decaying
to charged daughters only is required to lie within 15 \mevcc\
(2.5$\sigma$) of the $D^0$ mass, where $\sigma$ is the experimental
resolution.
The $D^0 \to K^- \pi^+ \pi^0$ candidates are required to have a mass
within 30 \mevcc\ (2.5$\sigma$) of the $D^0$ mass and to be located at a
point in the $D^0$ Dalitz plot where the density of events is larger than
1.4\% of the maximum density.\\
\indent
The $D^{*+}$ candidates are accepted if the mass difference
$m_{D^{*+}} - m_{D^0}$ is within 2 \mevcc\ (3$\sigma$) of the nominal
value, except for the $D^0 \to K^- \pi^+ \pi^0$ candidates where we use
1.5 \mevcc\ to reduce this mode's larger combinatoric background.

We combine oppositely-charged tracks from a common vertex into $\KS$
candidates. The \KS candidates are required to have a mass within 7
\mevcc\ (3$\sigma$) of the $\KS$ mass and a transverse flight length that
is significantly (4$\sigma$) greater than zero.

To form $B^0$ candidates, the $D^{(*)+}$ candidates are combined with
a $\KS$ candidate and a $\pi^-$, for which the particle identification
(PID) is inconsistent with being a kaon or an electron.
The probability of a common vertex is required to be above 0.1\%.
Using the beam energy, two almost-independent kinematic variables are
constructed:
the beam-energy substituted mass $\mes  \equiv \sqrt{(\sqrt{s}/2)^2
-{p^*_B}^2}$, and the difference between the $B^0$ candidate's measured
energy and the beam energy, $\de \equiv E^*_B - \sqrt{s}/2$. The asterisk
denotes evaluation in the \FourS\ CM frame.
$B^0$ candidates are required to have \de\ in the range $[-0.1,0.1]$ \gev,
and \mes\ in the range $[5.24,5.29]$ ($[5.20,5.288]$) \gevcc\ for \dkpi
(\dskpi).

To suppress the dominant continuum background events, which have a more
jet-like shape than $B \Bbar$ events, we use a linear combination, \fis, of
four variables: $L_0=\sum_{i} p_i$, $L_2 =\sum_{i} p_i  |\cos \theta_i|^2$, 
and the absolute values of the cosine of the polar angles of the $B$
momentum and of the $B$ thrust direction \cite{thrust}.
Here, $p_i$ is the momentum and $\theta_i$ is the angle with respect to
the thrust axis of the signal $B$ candidate of the tracks and clusters not
used to reconstruct the $B$. All of these variables are calculated in the
CM frame.
The coefficients are chosen to maximize the separation between signal
Monte Carlo and 9.6 \invfb\ of continuum events from data taken 40 \mev\
below the $\Upsilon(4S)$ resonance (off-resonance data). \fis\ has
negligible correlations with \mes\ and \de.

After the event selection, approximately 5\% of the events have more
than one $B^0$ candidate. We choose the one with $m_D$ closest to
the expected value and correct for differences between data and
simulation.
In simulated signal events, the final selection is 19.3\% efficient for
\bdkpi\ and 15.5\%, 3.9\% and 8.2\% efficient for \bdskpi\ in the three
$D^0$ decay modes $K^- \pi^+$, $K^- \pi^+ \pi^0$ and $K^- \pi^+ \pi^-
\pi^+$, respectively.

We perform an unbinned extended maximum likelihood fit with the variables
\mes, \de, and \fis\ on the selected candidates, using the logarithm of
the likelihood:
\begin{eqnarray}
  \ln \mathcal{L} = \!\!\!
    \sum_{\mbox{\tiny $i$=events}} \ln
      \left( \sum_{\mbox{\tiny $j$}} N_j P_{ij}(\mes,\de,\fis) \right) -
    \sum_{\mbox{\tiny $j$}} N_j,
\label{eq:llh}
\end{eqnarray}
where $P_{ij}$ is the product of probability density functions (PDFs) for
event $i$ of \mes, \de, and \fis, and $N_{j}$ is the number of events of
each sample component $j$: signal, continuum, combinatoric $B \Bbar$ decays,
and $B \Bbar$ events that peak in \mes\ but not in \de\ signal region
(denoted peaking $B \Bbar$ background).

The signal is described by a Gaussian distribution in \mes, two Gaussian
distributions with common mean in \de, and a Gaussian distribution with
different widths on each side of the mean (``bifurcated Gaussian
distribution'') in \fis. Their shape is obtained from the high-statistics
data control samples $B^0 \to D^{(*)\mp} a_1^\pm$ (similar topology of the
final state as the signal) for \mes\ and \de, and $B^0 \to D^{*\mp}
\pi^\pm$ for \fis, and all nine parameters are fixed in the fit.\\
\indent
The continuum and combinatoric $B \Bbar$ backgrounds are described
by empirical endpoint functions \cite{argus} in \mes, linear functions in
\de, and bifurcated Gaussian distributions in \fis. The \fis\ distribution
of continuum is obtained from off-resonance data, while the \fis\
distribution of the $B \Bbar$ backgrounds is obtained from Monte Carlo
simulation, and compared with data in high-statistics samples to ensure
that there is no significant difference.
The two \fis\ distributions and the common endpoint in \mes\ are fixed in
the fit, while the \mes\ shape and \de\ distributions are left free to
float, leaving four out of eleven parameters free in the fit.\\
\indent
The peaking $B \Bbar$ background is parametrized by a Gaussian
distribution in \mes, an exponential distribution in \de, and shares the
PDF in \fis\ with the non-peaking $B \Bbar$ background. The mean and width
in \mes\ of the peaking $B \Bbar$ background are fixed to values obtained
from Monte Carlo simulation, which are consistent with values measured in
\de\ sideband of data, thus adding one free and two fixed parameters.

The likelihood function is determined by the 27 parameters described
above, of which all four yields and five background shape parameters are
fitted. Subsequent to the fit, possible residual backgrounds from
combinatoric $D$ and \KS\ candidates are estimated using the sidebands of
$m_D$ and $m_{\KS}$, and subtracted.

The three-body and quasi-two-body (that is \bdpskstar) branching
fractions are obtained by fitting first without regard to event positions
in the Dalitz plot, and then with the requirement that the $\KS \pi^+$
invariant mass lies within 100 \mevcc\ of the $K^{*+}(892)$ mass.
Due to the relatively small number of background events in the second fit,
all $B \Bbar$ shape parameters are kept fixed to the value obtained in the
first fit.
\begin{figure}[t!]
\begin{center}
\setlength{\unitlength}{1mm}
\begin{picture}(80,170)(0,0)
  \jput(-10,115){\epsfig{file=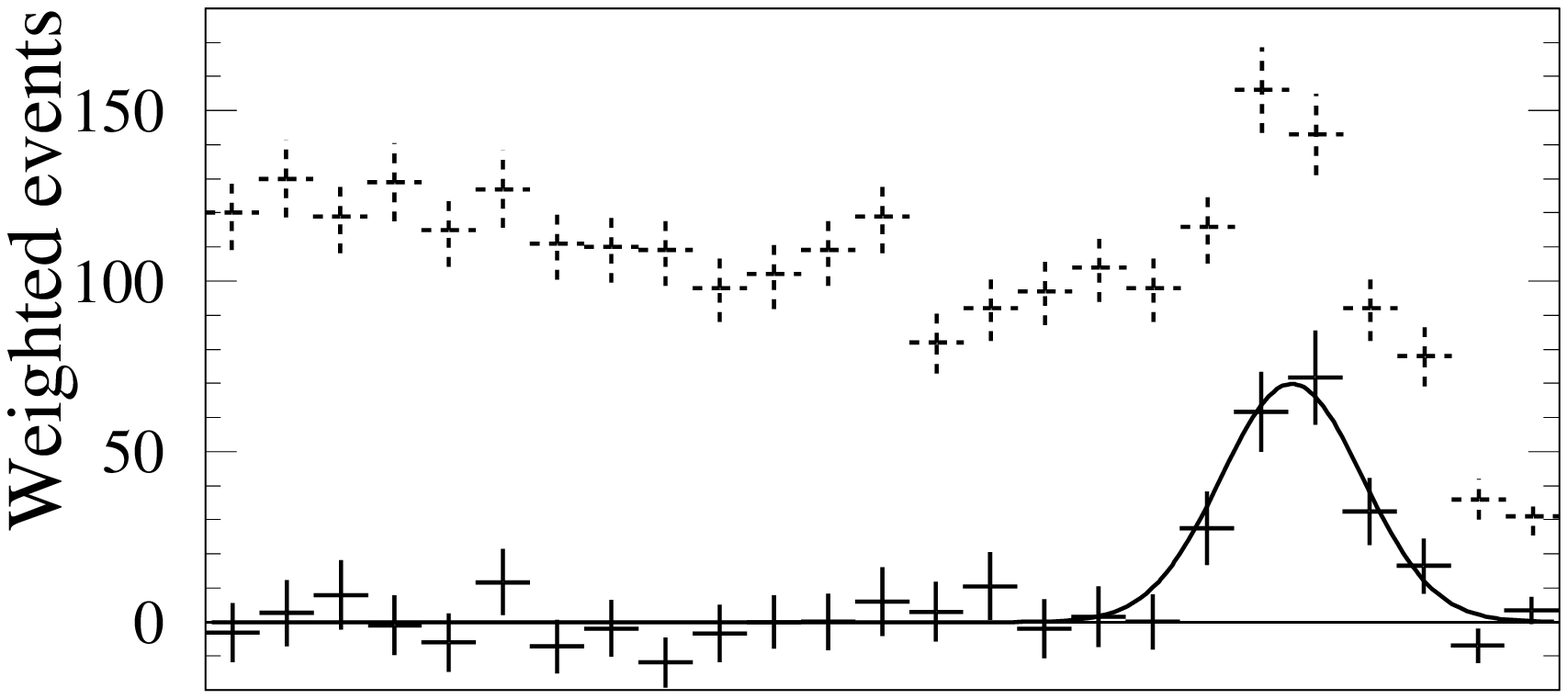,width=100mm}}
  \jput(15,163){\bdkpi}
  \jput(-10,77){\epsfig{file=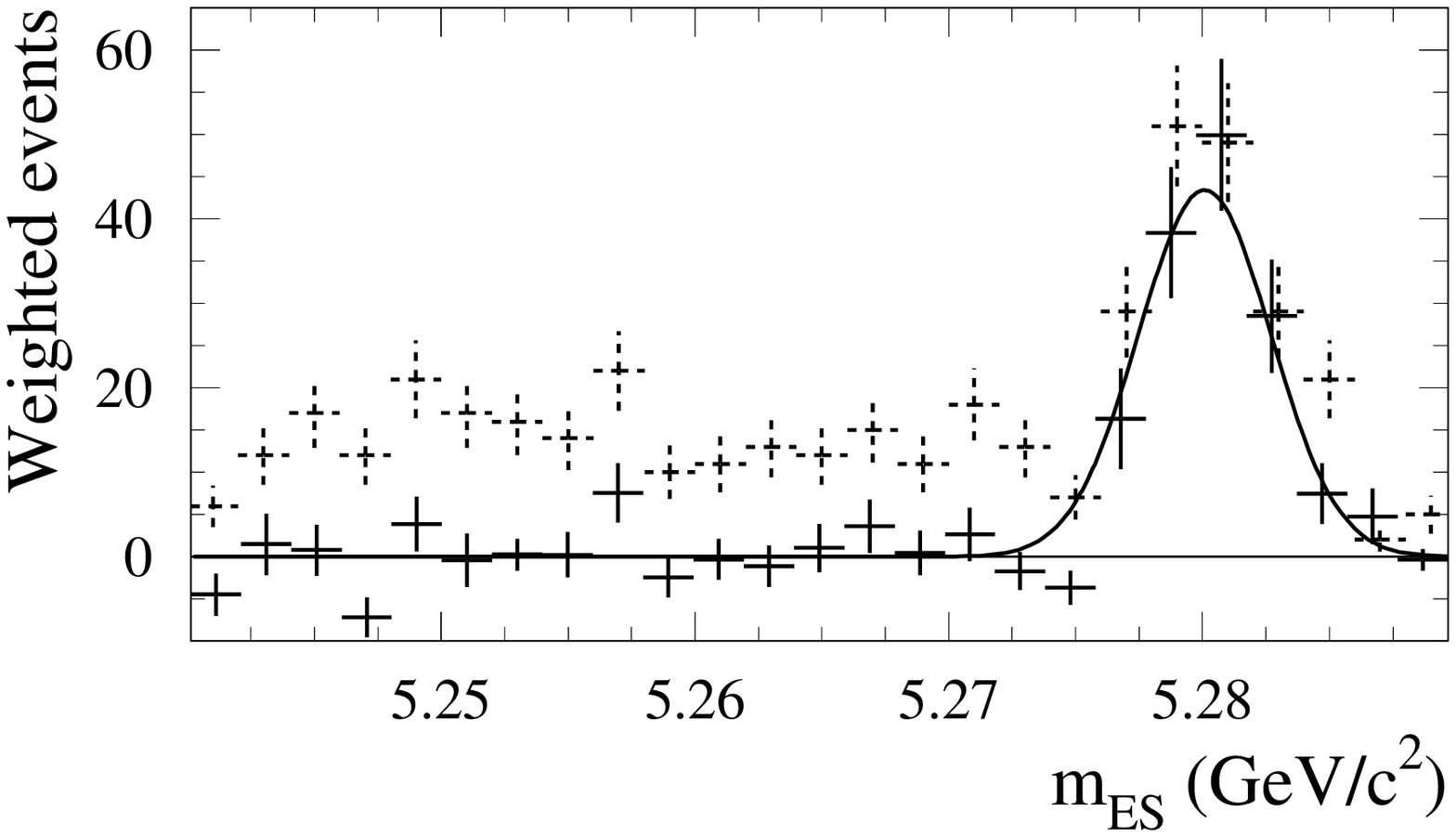,width=100mm}}
  \jput(15,125){\bdkstar}
  \jput(-10,26){\epsfig{file=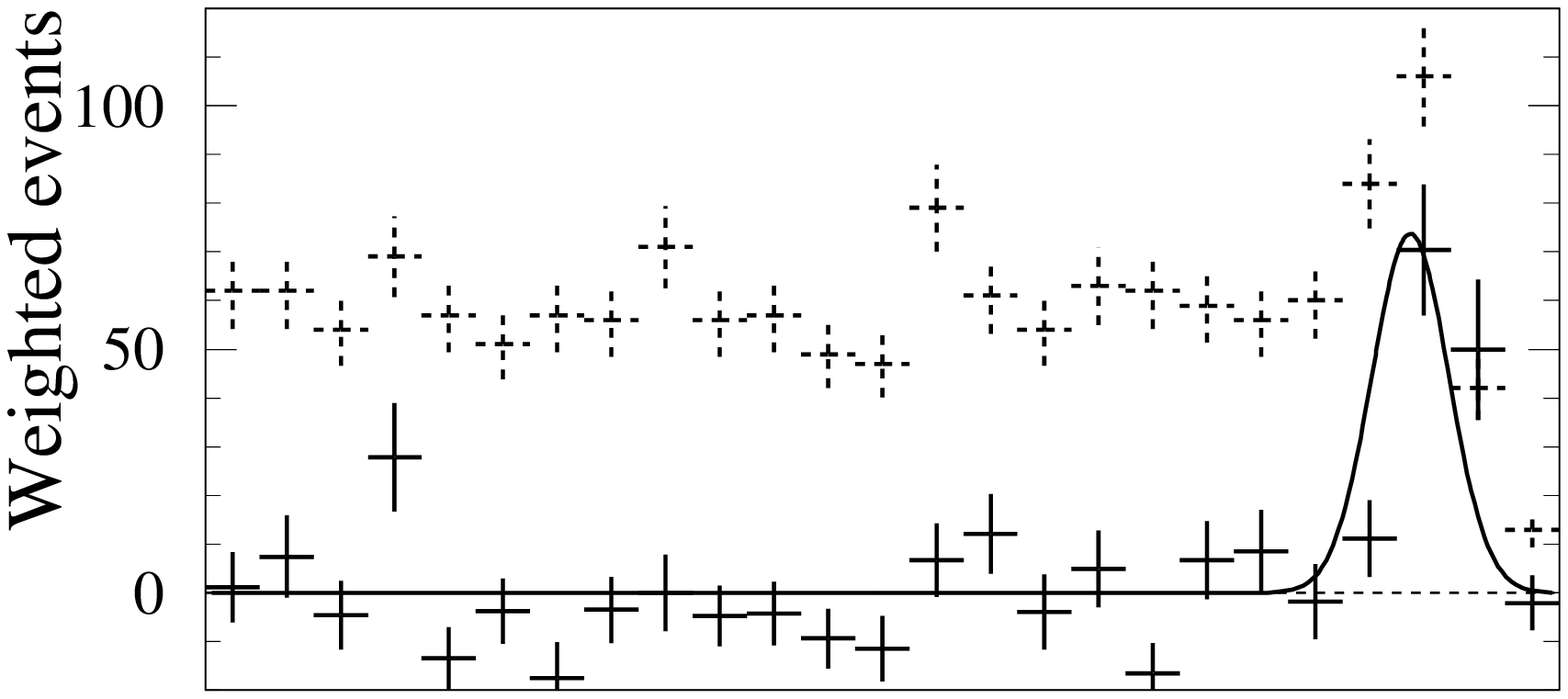,width=100mm}}
  \jput(15,74){\bdskpi}
  \jput(-10,-12){\epsfig{file=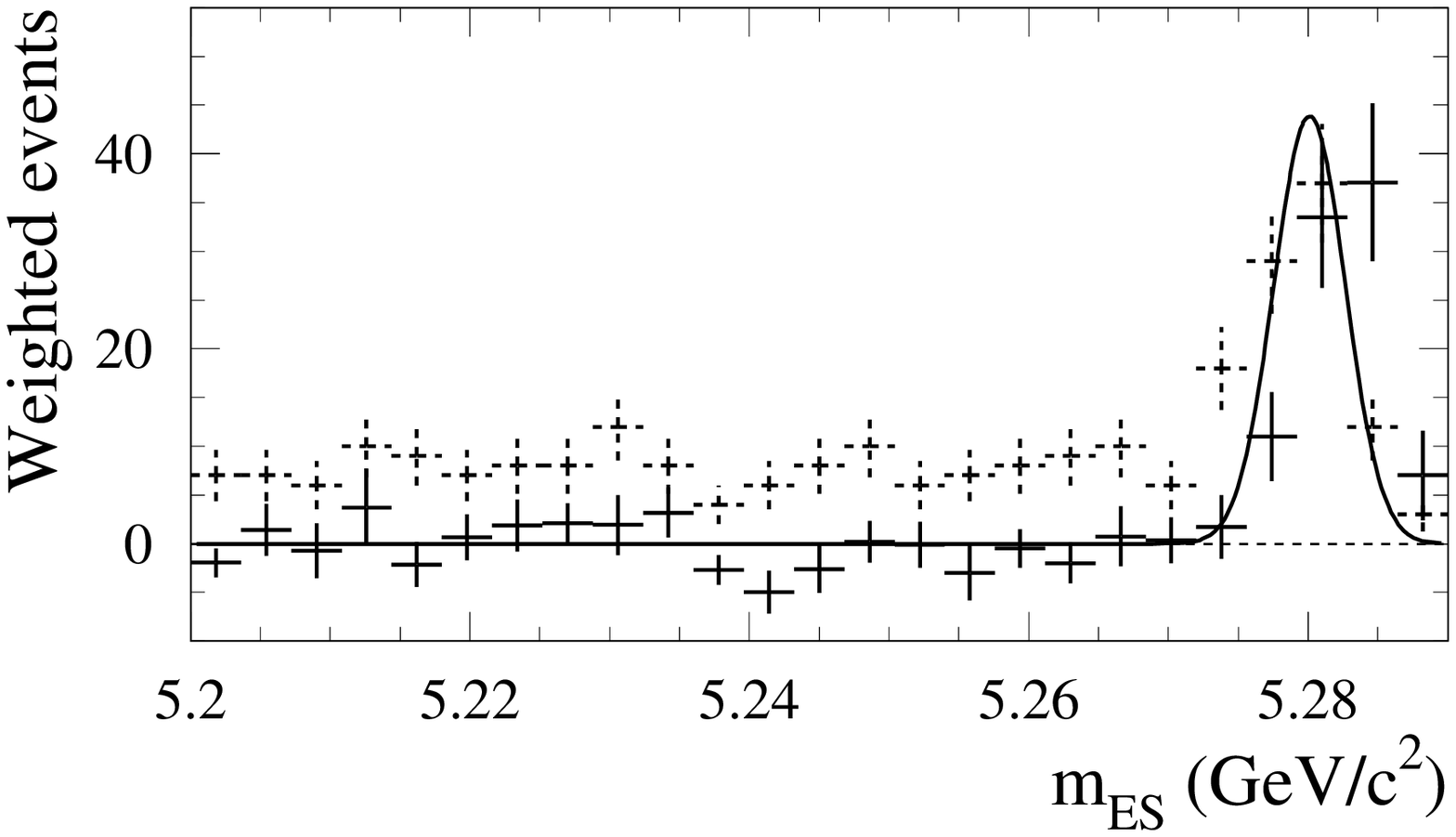,width=100mm}}
  \jput(15,36){\bdskstar}
\end{picture}
\vskip3mm
\caption{\mes distributions in data for the four decay modes. Events
  appropriately weighted by $W_{\mbox{\tiny sig}}$ (see text) to exhibit
  the signal distribution \cite{splot} are shown as solid points over
  which the fitted signal PDF is superimposed. For comparison, the \mes
  distribution obtained with $|\Delta E| < 25\mev$ (2$\sigma$) is included
  (dotted points).}
\label{fig:mes}
\end{center}
\vspace*{-2ex}
\end{figure}

The results are shown in Fig.\ \ref{fig:mes}, while yields and purities
(defined as $N_{\mbox{\tiny sig}}/\sigma^2(N_{\mbox{\tiny sig}})$)
are listed in Table \ref{tab:yields}, with the $K^{*+}$ resonant part
included in the three-body state.
\begin{table}[htbp]
\begin{center}
\leavevmode
\caption{\mbox{Signal yields and purities.}}
\begin{tabular}{lcc}
\hline
\hline
  Decay mode   &Signal yield        &Purity\\
\hline
  \bdkpi       &$230\pm24$          &$40\:\%$\\
  \bdkstar     &$143\pm14$          &$73\:\%$\\
  \bdskpi      &$134\pm17$          &$46\:\%$\\
  \bdskstar    &$ 78\pm10$          &$78\:\%$\\
\hline
\hline
\end{tabular}
\label{tab:yields}
\end{center}
\end{table}
\indent
To determine the three-body branching fractions optimally, a mapping of the
efficiency across the Dalitz plot is needed. This is obtained from simulated
signal events. Incorporating the efficiency variations ($\sim \pm 30\%$)
across the Dalitz plot requires a measure of the ({\it apriori} unknown)
event distribution in the Dalitz plot.
We obtain the number of signal events from the likelihood fit using
weights defined as:
\begin{eqnarray}
  W_{\mbox{\tiny sig}}^i &\equiv&
    \frac{\sum_j {\mathbf V}_{\mbox{\tiny sig},j} \: P_{ij}(\mes,\de,\fis)}
         {\sum_j N_j \: P_{ij}(\mes,\de,\fis)},
\end{eqnarray}
where $N_j$ and $P_{ij}$ are defined as in Eq.\ (\ref{eq:llh}), and
$\mathbf{V}_{\mbox{\tiny sig},j}$ is the signal row of the covariance
matrix of the component yields obtained from the likelihood fit.
These weights $W_{\mbox{\tiny sig}}^i$, which in the absence of correlations
are signal probabilities $P_{\mbox{\tiny sig}}/P_{\mbox{\tiny total}}$,
contain the signal distribution and its uncertainty for any quantity
uncorrelated with the variables in the likelihood fit \cite{splot}.

\begin{figure}[t!]
\begin{center}
\setlength{\unitlength}{1mm}
\begin{picture}(80,120)(0,0)
  \jput(-10,64){\epsfig{file=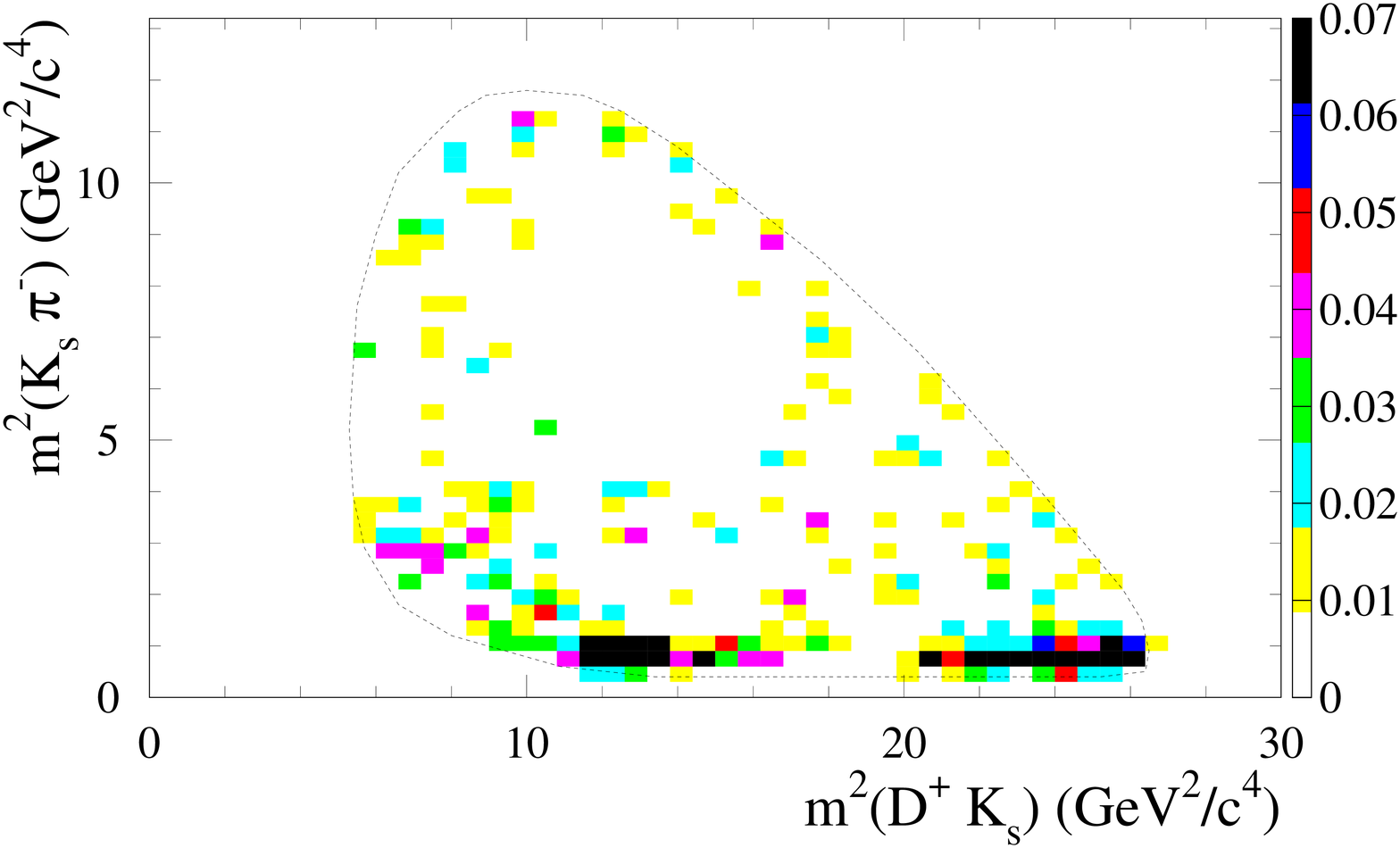,width=94mm}}
  \jput(48,115){\bdkpi}
  \jput(-10,10){\epsfig{file=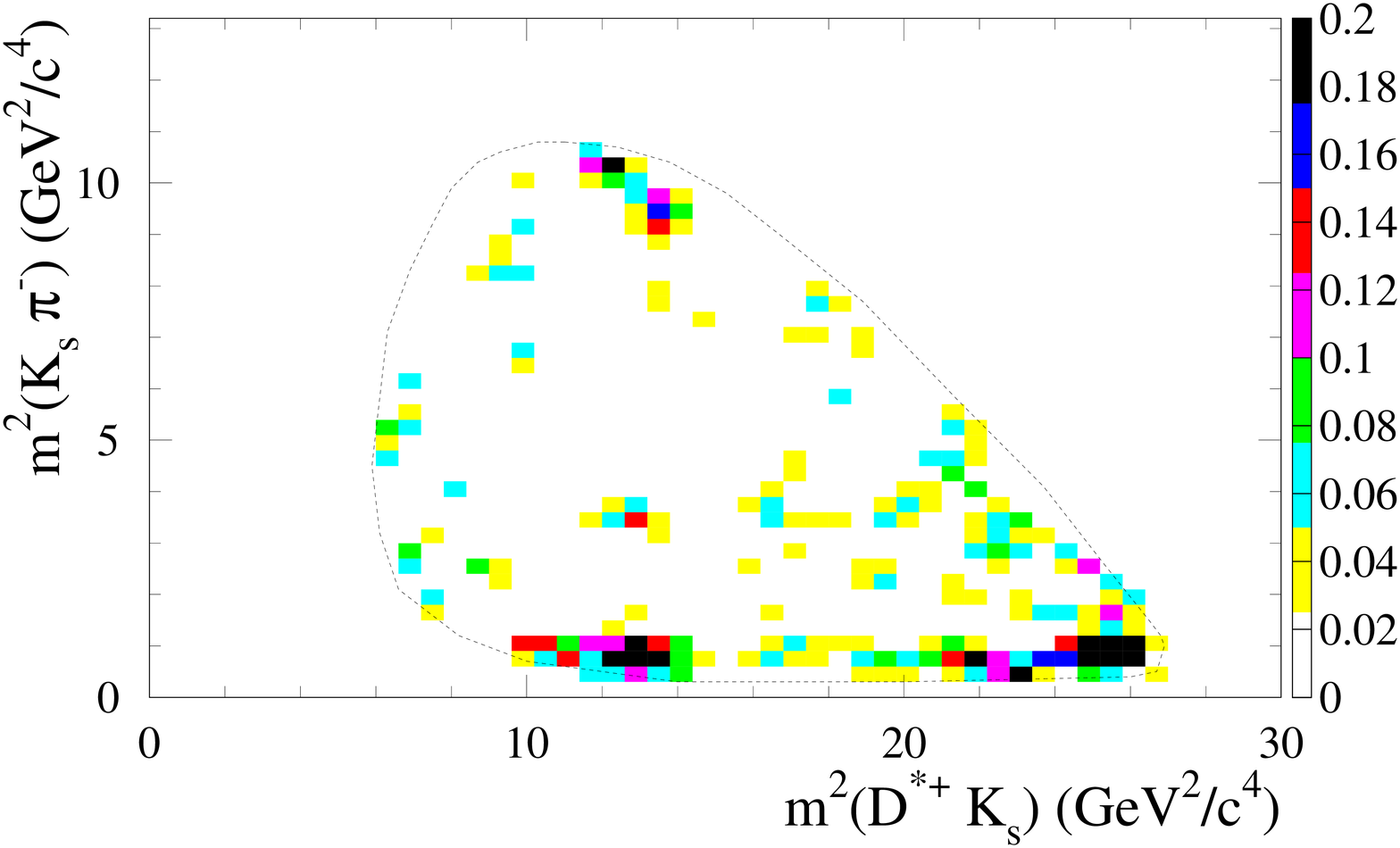,width=94mm}}
  \jput(48,61){\bdskpi}
\end{picture}
\vskip -17mm
\caption{Signal Dalitz distributions with events weighted by
  $W_{\mbox{\tiny sig}}$ and corrected for efficiency variations. Each bin
  is colored according to its contribution to the branching fraction. The
  bins in white also include the contributions which are negative but
  still statistically compatible with zero.}
\label{fig:dalitz}
\end{center}
\vspace*{-2ex}
\end{figure}

\begin{figure}[t!]
\begin{center}
\setlength{\unitlength}{1mm}
\begin{picture}(80,55)(0,0)
  \jput(-10,-5){\epsfig{file=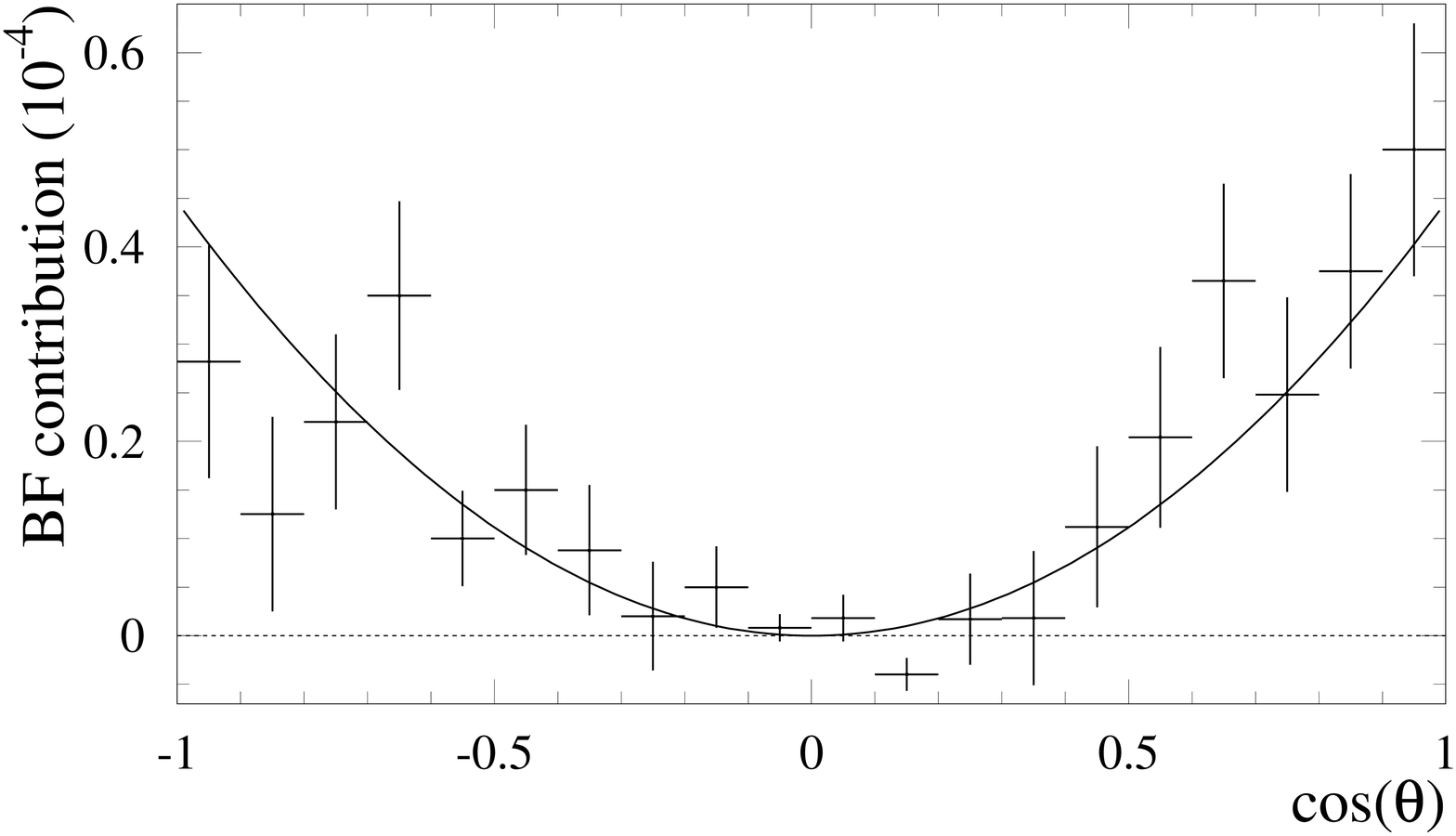,width=95mm}}
  \jput(10,45){\bdkstar}
\end{picture}
\vskip -5mm
\caption{Distribution of $\cos\theta$ for data for the \bdkpi\ decay mode
  in the $K^{*\pm}$ region using the signal weights $W_{\mbox{\tiny sig}}$
  and correcting for efficiency variations. The solid curve shows the
  expected spin-1 distribution $dN/d\cos\theta \propto \cos^2 \theta$.}
\label{fig:hel}
\end{center}
\vspace*{-2ex}
\end{figure}

The efficiency-corrected Dalitz distributions, weighted by
$W_{\mbox{\tiny sig}}$, are shown in Fig. \ref{fig:dalitz}.
The $K^*(892)^+$ resonance is dominant in both the \bdkpi\ and \bdskpi\
modes, while no other resonant structures are significant.
In the \bdkpi\ channel, the spin-1 $K^{*\pm}$ meson has the helicity
distribution $dN/d\cos\theta \propto \cos^2 \theta$, where $\theta$ is the
angle between the $K^{*\pm}$ and the $K^0$ in the $K^{*\pm}$ center of
mass frame. This can be seen in Fig.~\ref{fig:hel}.

%
\indent
The systematic errors are summarized in Table \ref{tab:systematics}. Most
systematic errors are due to possible differences between data and Monte
Carlo.
The tracking efficiency residuals and associated systematic error are
obtained from a large sample of $\tau$ decays.
The efficiency correction as a function of the position in the Dalitz plot
obtained from simulated signal events comes with systematic uncertainties
due to resolution effects and binning, which are mostly of statistical
origin.
A $\pm 1\sigma$ variation of all fixed variables in the fit, including
relevant correlations, is used to obtain the systematic from the
uncertainty in the PDFs.
\begin{table}[htbp]
\begin{center}
\leavevmode
\caption{Sources and sizes of systematic errors. The combined errors take
  correlations into account. All numbers are in percent.}
\begin{tabular}{lcccc}
\hline
\hline
  Systematic                   &$DK\pi$  &$D K^*$  &$D^*K\pi$  &$D^*K^*$\\
\hline
  Tracking efficiency          &$5.9$    &$5.9$    &$6.1$      &$6.3$\\
  PID efficiency               &$2.2$    &$2.0$    &$2.0$      &$2.0$\\
  ${\cal B}(D^{*+})$           &$ - $    &$ - $    &$0.7$      &$0.7$\\
  ${\cal B}(D^{+/0})$          &$6.5$    &$6.5$    &$3.4$      &$3.8$\\
  $D^{(*)}$ reconstruction     &$0.7$    &$0.7$    &$1.2$      &$1.2$\\
  $K^{*+}$ fraction fit        &$ - $    &$3.7$    &$ - $      &$5.1$\\    
  ${\cal B}(\KS)$              &$0.2$    &$0.2$    &$0.2$      &$0.2$\\
  \KS reconstruction           &$1.8$    &$1.9$    &$1.9$      &$1.9$\\
  $\pi^0$ reconstruction       &$ - $    &$ - $    &$0.8$      &$1.2$\\
  PDF parametrization          &$4.5$    &$2.9$    &$7.1$      &$3.7$\\
  Efficiency variation         &$3.5$    &$4.9$    &$6.3$      &$5.6$\\
  $B \Bbar$ counting           &$1.1$    &$1.1$    &$1.1$      &$1.1$\\
\hline
  Combined error               &$11.0$   &$11.6$   &$12.6$     &$12.2$\\
\hline
\hline
\end{tabular}
\label{tab:systematics}
\end{center}
\end{table}

\indent
Our final branching fraction results, weighting the three $D^0$ modes
according to their combined statistical and uncorrelated systematic error,
are:
\begin{eqnarray*}
  {\cal B}(\bdkpi)    \!\! &=& \!\! 
  (4.9 \pm 0.7_{\, \mbox{\footnotesize stat}}   
        \pm 0.5_{\, \mbox{\footnotesize syst}}) \! \times \! 10^{-4},\\
  {\cal B}(\bdskpi)   \!\! &=& \!\! 
  (3.0 \pm 0.7_{\, \mbox{\footnotesize stat}}   
        \pm 0.3_{\, \mbox{\footnotesize syst}}) \! \times \! 10^{-4},\\
  {\cal B}(\bdkstar)  \!\! &=& \!\! 
  (4.6 \pm 0.6_{\, \mbox{\footnotesize stat}}   
        \pm 0.5_{\, \mbox{\footnotesize syst}}) \! \times \! 10^{-4},\\
  {\cal B}(\bdskstar) \!\! &=& \!\! 
  (3.2 \pm 0.6_{\, \mbox{\footnotesize stat}}   
        \pm 0.3_{\, \mbox{\footnotesize syst}}) \! \times \! 10^{-4}.
\label{eq:results}
\end{eqnarray*}

To summarize, a clear signal is seen in both the \bdkpi\ and \bdskpi\
channels, and in both modes the $K^*(892)^+$ resonance is dominant.
Defining the $K^*$ resonant fractions, $f$, as 
${\cal B}(\bdpskstar) {\cal B}(K^{*+} \to K^0 \pi^+) / {\cal B}(\bdpskpi)$, 
we obtain the fractions
$f(\bdkstar)  = 0.63 \pm 0.08_{\,\mbox{\footnotesize stat}}
                     \pm 0.04_{\,\mbox{\footnotesize syst}}$ 
and
$f(\bdskstar) = 0.72 \pm 0.14_{\,\mbox{\footnotesize stat}}
                     \pm 0.05_{\,\mbox{\footnotesize syst}}$,
                     respectively,
where the systematic errors are mainly from correcting for any possible
non-resonant contributions.\\
Both the method of this analysis and the resulting three-body branching
fraction measurements are the first of their kind, while the resonant
decay modes have been measured before \cite{cleo}.
To determine the sensitivity to $\gamma$ of these modes, a time-dependent
Dalitz fit is required, for which the data sample is inadequate. However,
the branching fractions and Dalitz distributions suggest that these modes
will be useful for measuring $\gamma$ at the $B$-factories.


We are grateful for the excellent luminosity and machine conditions
provided by our \pep2\ colleagues, 
and for the substantial dedicated effort from
the computing organizations that support \babar.
The collaborating institutions wish to thank 
SLAC for its support and kind hospitality. 
This work is supported by
DOE
and NSF (USA),
NSERC (Canada),
IHEP (China),
CEA and
CNRS-IN2P3
(France),
BMBF and DFG
(Germany),
INFN (Italy),
FOM (The Netherlands),
NFR (Norway),
MIST (Russia), and
PPARC (United Kingdom). 
Individuals have received support from the 
A.~P.~Sloan Foundation, 
Research Corporation,
and Alexander von Humboldt Foundation.


\end{document}